\documentclass{aa}

\usepackage{graphicx}
\usepackage[varg]{txfonts}
\usepackage{natbib}
\bibpunct{(}{)}{;}{a}{}{,}
\usepackage{array}
\usepackage{xspace}
\usepackage{lscape}
\usepackage{url}
\usepackage{arydshln}

\usepackage[hyperfootnotes=false, linktocpage=true, breaklinks=false, colorlinks=true, linkcolor=blue, citecolor=blue, urlcolor=blue]{hyperref}
\usepackage[all]{hypcap}

\newcommand{\FeKa}{Fe K\ensuremath{\alpha}\xspace}

\newcommand{\kms}{\ensuremath{\mathrm{km\ s^{-1}}}\xspace}
\newcommand{\NH}{\ensuremath{N_{\mathrm{H}}}\xspace}

\newcommand{\xmm}{{\it XMM-Newton}\xspace}
\newcommand{\chandra}{{\it Chandra}\xspace}
\newcommand{\swift}{{\it Swift}\xspace}

\newcommand{\integral}{{INTEGRAL}\xspace}
\newcommand{\hst}{{HST}\xspace}

\newcommand{\ngc}{{NGC~5548}\xspace}

\newcommand{\nustar}{{\it NuSTAR}\xspace}
\newcommand{\fvar}{\ensuremath{F_{\rm{var}}}\xspace}
\newcommand{\cf}{\ensuremath{C_f}\xspace}

\newcommand{\ergcm}{{\ensuremath{\rm{erg\ cm}^{-2}\ \rm{s}^{-1}\ \mbox{\AA}^{-1}}}\xspace}

\defcitealias{Meh14a}{Paper I}
\defcitealias{Kaas14}{K14}

\begin{document}

\title{Anatomy of the AGN in NGC 5548}

\subtitle{VII. {\textbf{\emph{Swift}}} study of obscuration and broadband continuum variability}

\author{
M. Mehdipour \inst{1,2} 
\and
J.S. Kaastra \inst{1,3,4}
\and 
G.A. Kriss \inst{5,6}
\and
M. Cappi \inst{7}
\and
P.-O. Petrucci \inst{8,9}
\and
B. De Marco \inst{10}
\and
G. Ponti \inst{10}
\and
K.C. Steenbrugge \inst{11}
\and
E. Behar \inst{12}
\and 
S. Bianchi \inst{13}
\and
G. Branduardi-Raymont \inst{2}
\and
E. Costantini \inst{1}
\and
J. Ebrero \inst{14,1}
\and
L. Di Gesu \inst{1}
\and 
G. Matt \inst{13}
\and
S. Paltani \inst{15}
\and
B.M. Peterson \inst{16,17}
\and
F. Ursini \inst{8,9,13}
\and
M. Whewell \inst{2}
}

\institute{
SRON Netherlands Institute for Space Research, Sorbonnelaan 2, 3584 CA Utrecht, the Netherlands\\ \email{M.Mehdipour@sron.nl}
\and
Mullard Space Science Laboratory, University College London, Holmbury St. Mary, Dorking, Surrey, RH5 6NT, UK
\and
Department of Physics and Astronomy, Universiteit Utrecht, P.O. Box 80000, 3508 TA Utrecht, the Netherlands
\and
Leiden Observatory, Leiden University, PO Box 9513, 2300 RA Leiden, the Netherlands
\and
Space Telescope Science Institute, 3700 San Martin Drive, Baltimore, MD 21218, USA
\and
Department of Physics and Astronomy, The Johns Hopkins University, Baltimore, MD 21218, USA
\and
INAF-IASF Bologna, Via Gobetti 101, I-40129 Bologna, Italy
\and
Univ. Grenoble Alpes, IPAG, F-38000 Grenoble, France
\and
CNRS, IPAG, F-38000 Grenoble, France
\and
Max-Planck-Institut f\"{u}r extraterrestrische Physik, Giessenbachstrasse, D-85748 Garching, Germany
\and
Instituto de Astronom\'{i}a, Universidad Cat\'{o}lica del Norte, Avenida Angamos 0610, Casilla 1280, Antofagasta, Chile
\and
Department of Physics, Technion-Israel Institute of Technology, 32000 Haifa, Israel
\and
Dipartimento di Matematica e Fisica, Universit\`{a} degli Studi Roma Tre, via della Vasca Navale 84, 00146 Roma, Italy
\and
European Space Astronomy Centre, P.O. Box 78, E-28691 Villanueva de la Ca\~{n}ada, Madrid, Spain
\and
Department of Astronomy, University of Geneva, 16 Ch. d'Ecogia, 1290 Versoix, Switzerland
\and
Department of Astronomy, The Ohio State University, 140 W 18th Avenue, Columbus, OH 43210, USA
\and
Center for Cosmology \& AstroParticle Physics, The Ohio State University, 191 West Woodruff Ave., Columbus, OH 43210, USA
}

\date{Received 11 November 2015 / Accepted 6 February 2016}

\abstract
{We present our investigation into the long-term variability of the X-ray obscuration and optical-UV-X-ray continuum in the Seyfert~1 galaxy \ngc. In 2013 and 2014, the \swift observatory monitored \ngc on average every day or two, with archival observations reaching back to 2005, totalling about 670~ks of observing time. Both broadband spectral modelling and temporal rms variability analysis are applied to the \swift data. We disentangle the variability caused by absorption, due to an obscuring weakly-ionised outflow near the disk, from variability of the intrinsic continuum components (the soft X-ray excess and the power-law) originating from the disk and its associated coronae. The spectral model that we apply to this extensive \swift data is the global model that we derived for \ngc from analysis of the stacked spectra from our multi-satellite campaign of summer 2013 (including \xmm, \nustar and HST). The results of our \swift study show that changes in the covering fraction of the obscurer is the primary and dominant cause of variability in the soft X-ray band on timescales of 10 days to ${\sim 5}$ months. The obscuring covering fraction of the X-ray source is found to range between 0.7 and nearly 1.0. The contribution of the soft excess component to the X-ray variability is often much less than that of the obscurer, but it becomes comparable when the optical-UV continuum flares up. We find that the soft excess is consistent with being the high-energy tail of the optical-UV continuum and can be explained by warm Comptonisation: up-scattering of the disk seed photons in a warm, optically thick corona as part of the inner disk. To this date, the \swift monitoring of \ngc shows that the obscurer has been continuously present in our line of sight for at least 4 years (since at least February 2012).}

\authorrunning{M. Mehdipour et al.}
\titlerunning{Anatomy of the AGN in NGC 5548. VII.}
\maketitle

\section{Introduction}

The growth of supermassive black holes (SMBHs) in active galactic nuclei (AGN) is accompanied by powerful jets and/or winds of ionised gas. However, the association between the accretion and outflow phenomena in AGN is poorly understood, leading to important and outstanding questions in AGN science (see e.g. the review by \citealt{Fabian12}). Some of these questions are: which physical parameters regulate accretion and outflows; how is the energy budget distributed between accretion, outflows and radiative output, and how are they dependent on the SMBH mass and accretion rate; what are the implications of the outflows on their host galaxies and beyond?

As ionised outflows are ultimately powered and driven by energy released from the accretion process (e.g. \citealt{Prog00}), their properties are expected to be related to the physical conditions and radiation spectra of the accretion disk and its associated higher-energy components. Determining the physical and ionisation structure, dynamics and origin of the ionised outflows, as well as their role in shaping AGN spectra and variability, are crucial requirements for advancing our knowledge of AGN. However, there are challenges before any global characterisation of the ionised outflows, their link to accretion, and their impact on their environment can be established. One difficulty hampering our understanding of outflows is the uncertain nature of the observed X-ray spectral variability. For example, both accretion-powered continuum emission and absorption by outflows may contribute to the observed spectral variability, and their disentanglement is challenging. Moreover, the origin of different intrinsic broadband components, and their associations with each other, are not fully understood due to convolution of these components across wide energy bands. Monitoring the spectral variability of AGN at optical-UV-X-ray energies and incorporating both high-resolution spectroscopy and temporal analysis techniques is the most effective way of overcoming this challenge. Our extensive multi-wavelength campaign on the Seyfert~1 galaxy \object{NGC~5548}, introduced in \citet{Meh14a} (hereafter \citetalias{Meh14a}), provides a rare opportunity to determine the nature and origin of spectral variability in AGN and understand the physical processes which give rise to the formation of the AGN spectral energy distribution (SED).

From our campaign on \ngc, an exceptional kind of X-ray obscuring outflow was discovered at the core of this AGN by \citet{Kaas14} (hereafter \citetalias{Kaas14}). In contrast to the commonly seen ionised winds at pc-scale distances from the SMBH (e.g. \object{Mrk~509}, \citealt{Kaas12}), the remarkable obscurer in this AGN is a new breed of weakly-ionised, yet high-velocity (${\sim 5000~\kms}$) outflowing wind close to the SMBH at distances of a few light days. It also extends into the broad-line region (BLR). This obscurer most likely originates from the accretion disk. Such obscuring disk winds have important implications for the launch of outflows and feedback dynamics in AGN. The obscuration shields gas from X-ray and extreme ultraviolet (EUV) radiation at the starting point of the wind, which is essential in order to drive away the wind using UV radiation in quasars (e.g. \citealt{Prog04}). Bright and nearby Seyfert galaxies such as \ngc are ideal laboratories for studying the mechanisms that drive powerful winds in the more luminous quasars (which are too faint in X-rays for a detailed spectral analysis), and can regulate the growth and co-evolution of SMBHs and their host galaxies.

The obscurer in \ngc is partially covering the central X-ray emitting source (\citetalias{Kaas14}). Continuous and frequent monitoring with \swift over a few years enables us to trace the variability at optical-UV-X-ray energies, which is used to derive a precise characterisation of the continuum and obscuration variability. In order to obtain the continuum variability as far as possible into the UV energies, we make use of UV data from a Hubble Space Telescope (HST) Cosmic Origins Spectrograph (COS - \citealt{Green12}) monitoring program \citep{DeRosa15}. Using the obscuration model of \citetalias{Kaas14} and the broadband spectral model of \citetalias{Meh14a} (derived from stacked spectra of our campaign, including \xmm and \nustar), we perform a broadband spectral and timing analysis of the \swift and HST COS data. This enables us to disentangle the X-ray variability caused by absorption (due to the obscurer) from different X-ray emission components originating from the accretion disk/coronal regions (the soft X-ray excess and the primary power-law), and study the nature and long-term variability of the obscuration and the optical-UV-X-ray continuum in \ngc.

The structure of this paper is as follows. Section \ref{obs_sect} describes the \swift monitoring of \ngc, with an overview of the lightcurves and spectra. In Sect. \ref{broadband_sect} we present our broadband spectral modelling and examine the derived variability of the model parameters. In Sect. \ref{RMS_sect} we present our rms variability analysis of the X-ray and optical/UV lightcurves at different energies, and compare the observed rms spectra with the results of our broadband spectral modelling. We discuss all our findings in Sect. \ref{discussion} and give concluding remarks in Sect. \ref{conclusions}. In Appendix \ref{uv_appendix} we show the relations between the \swift and HST COS optical/UV lightcurves at different energies.  

The spectral analysis and modelling presented in this paper were done using the {\tt SPEX}\,\footnote{\url{http://www.sron.nl/spex}} package \citep{Kaa96} v2.06.01. We also made use of tools in NASA's {\tt HEASOFT}\,\footnote{\url{http://heasarc.nasa.gov/lheasoft}} v6.14 package. The spectra shown in this paper are background-subtracted and are displayed in the observed frame. We use C-statistics for spectral fitting and give errors at $1\sigma$ (68\%) confidence level. The redshift of \ngc is set to 0.017175 \citep{deVa91}. The adopted cosmological parameters for distance and luminosity computations in our modelling are ${H_{0}=70\ \mathrm{km\ s^{-1}\ Mpc^{-1}}}$, $\Omega_{\Lambda}=0.70$ and $\Omega_{m}=0.30$.

\section{{\textbf{\emph{Swift}}} monitoring of \ngc}
\label{obs_sect}

\swift \citep{Gehr04} has been extensively and frequently monitoring \ngc over recent years, thanks to various observing programs. On average there was an observation about every week in 2012, every two days in 2013 and every day in the first half of 2014. Apart from these data taken during the obscured epoch of \ngc, there were also a few observations taken in April--May 2005 and June--August 2007 when the source was unobscured. There were no \swift observations between August 2007 and February 2012. The total \swift observing time of \ngc is 670 ks (up to 4 February 2015), of which 16 ks were taken in 2005, 21 ks in 2007, 62 ks in 2012, 327 ks in 2013, 211 ks in 2014 and 33 ks in 2015. The length of a \swift observation is 1--2 ks. In this study we have used all the \swift data taken up to 4 February 2015, when the regular \swift monitorings of \ngc ended. The details of reduction and processing of the \swift data from the X-ray Telescope (XRT - \citealt{Burr05}) and the UV/Optical Telescope (UVOT - \citealt{Romi05}), as well as other data from our campaign and also the observation logs, are presented in \citetalias{Meh14a}. 

The UVOT observations of \ngc have been mostly taken with the UVW2 filter in addition to other photometric filters. In a total of $668$ UVOT observations recorded up to 4 February 2015, 90\% of them had exposures taken with the UVW2 filter, 57\% with UVW1, 56\% with V, 52\% with each B and U, and 50\% with UVM2. This difference in coverage by the filters is due to various observing modes requested for UVOT and the filter-of-the-day selection at the time of the UVOT observations. The UVW2 filter provides the UV flux at shorter wavelengths (2030 \AA) than other filters and demonstrates the clearest and largest variability seen by the instrument. Hence, it is useful for simultaneous timing analysis with the X-rays from XRT. During most of the 2013 and 2014 monitorings, all six primary UVOT filters were utilised. 

For the purpose of determining the UV continuum variability at shorter wavelengths than that of UVOT, we used HST COS data of \ngc. The HST COS continuum data were incorporated together with \swift UVOT and XRT data in our broadband spectral modelling and rms variability analysis. We therefore selected only those COS observations which had contemporaneous \swift observations (separated by less than 24 hours). The contemporaneous \swift and HST COS observations of \ngc cover 120 days in total, 5 of which are from the multi-satellite campaign of 2013 (\citetalias{Meh14a}) and the remaining 115 days are from an optical/UV reverberation-mapping program carried out in 2014 \citep{DeRosa15}. The COS data consist of the observed continuum flux from five narrow wavelength bands (central wavelengths of 1158 \AA, 1367 \AA, 1462 \AA, 1479 \AA\ and 1746 \AA), which are free of emission and absorption features over the wavelength range covered by the COS G130M and G160M gratings. More details about the optical/UV lightcurves (HST COS and UVOT) and their relations are given in Appendix \ref{uv_appendix}.

\subsection{Swift lightcurves and spectra}
\label{swift_lc_sect}

\begin{figure*}[!tbp]
\centering
\resizebox{0.97\hsize}{!}{\includegraphics[angle=0]{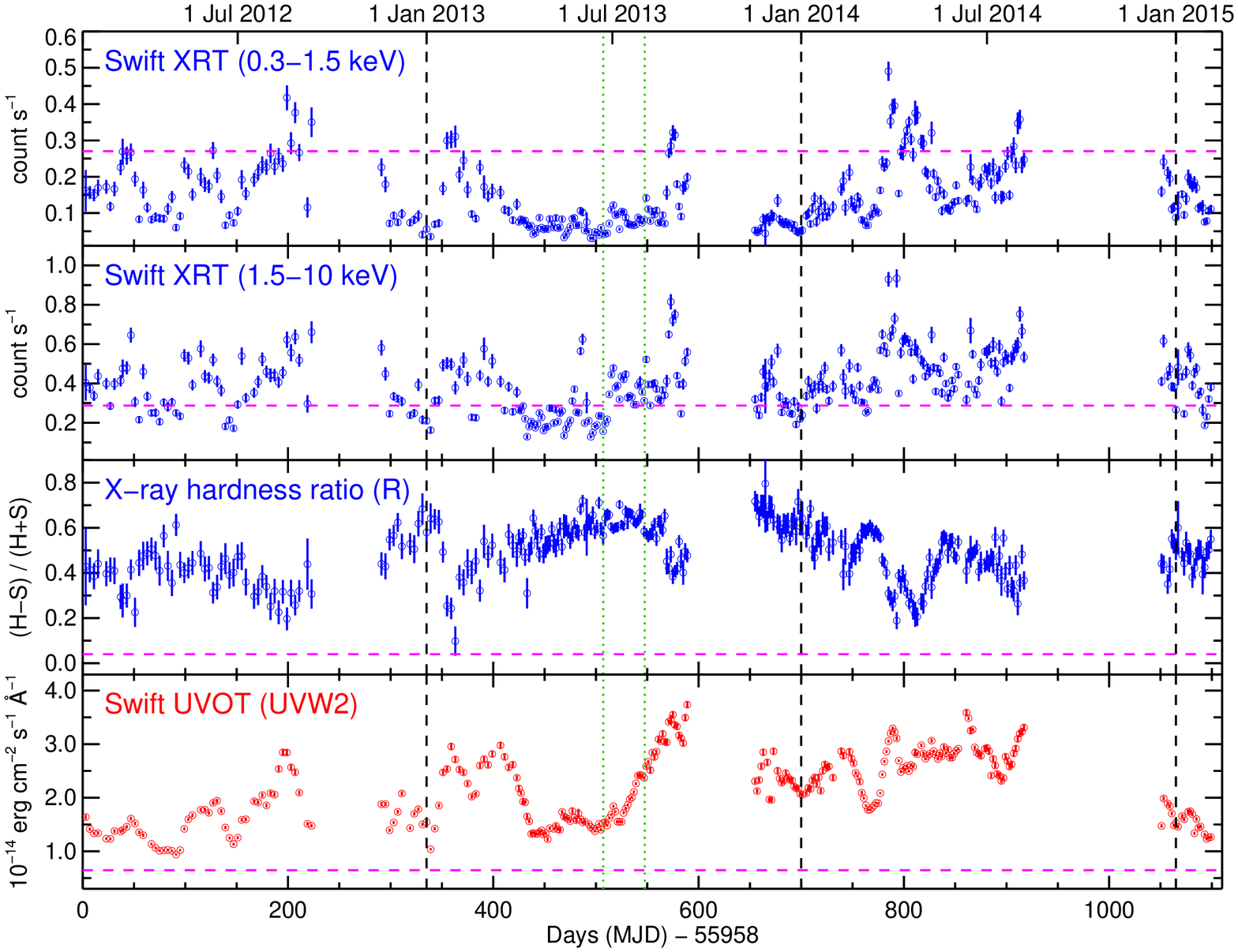}}
\caption{\swift lightcurves of \ngc from 1 February 2012 to 4 February 2015, during which the X-ray source is obscured. The data are displayed with bin sizes of two days for clarity of presentation. The vertical dotted lines in green indicate the interval of our summer 2013 \xmm campaign. The vertical dashed lines in black indicate the start of each year for reference. The horizontal dashed lines in magenta represent the averages from \swift observations in 2005 and 2007, when the X-ray source was unobscured. The panels are described in Sect. \ref{swift_lc_sect}.}
\label{swift_lc}
\end{figure*}

Figure \ref{swift_lc} shows the \swift lightcurves of \ngc from 1 February 2012 to 4 February 2015, covering a range of 1100 days during the obscured epoch. The figure displays the soft (0.3--1.5 keV) and hard (1.5--10 keV) X-ray count rate fluxes from XRT and the corresponding X-ray hardness ratio, as well as the observed UV flux in the UVW2 filter of UVOT. The X-ray hardness ratio ($R$) used here is defined as:
\begin{equation}
\label{hardness_eq}
R = (H-S) / (H+S)
\end{equation}
where $H$ and $S$ are the count rate fluxes in the hard (1.5--10 keV) and soft (0.3--1.5 keV) bands of XRT, respectively. 

The lightcurves of Fig. \ref{swift_lc} display interesting features that one notices by eye before any analysis. It is evident that most of the time, the soft X-ray flux is well below the average level of the unobscured epoch (indicated by the horizontal dashed line in magenta), whereas in the hard band the flux is at similar (or higher) levels to the unobscured epoch. As explained in \citetalias{Kaas14}, the suppression of the soft X-rays in \ngc is caused by an outflowing X-ray obscurer. The hardness ratio $R$, which is indicative of the soft X-ray absorption by the obscurer, shows the spectrum is continuously harder than during the unobscured epoch, when $R$ was about $0.04$. Thus, the \swift data indicate a long-lasting obscuration is occurring in \ngc. Furthermore, there appears to be a long-term gradual rise and decline of the hardness ratio in 2013 and 2014, pointing to a possible long-term variability of the obscurer. Interestingly, there is a significant positive correlation between the variability of the UV flux and that of the soft and hard X-ray bands (null hypothesis probability of $< 10^{-10}$).  Another feature is the anti-correlation between the X-ray hardness ratio and the UV flux: at some periods when the UV flux peaks, the hardness ratio dips (e.g. days 550--600 of the lightcurves), or when the UV dips, the hardness ratio peaks (e.g. days 740--790 of the lightcurves). This anti-correlation points to the soft X-ray excess component in \ngc being linked to the UV continuum (explained in Sect. \ref{para_results} and also \citetalias{Meh14a}), so that when the UV flux goes up, the source gets softer in the X-rays (i.e. the hardness ratio goes down). Following our spectral modelling in Sect. \ref{broadband_sect}, the correlations are examined in Sect. \ref{para_results}.

\begin{figure}[!tbp]
\centering
\vspace{-1.0cm}
\resizebox{1.11\hsize}{!}{\hspace{-1.1cm}\includegraphics[angle=0]{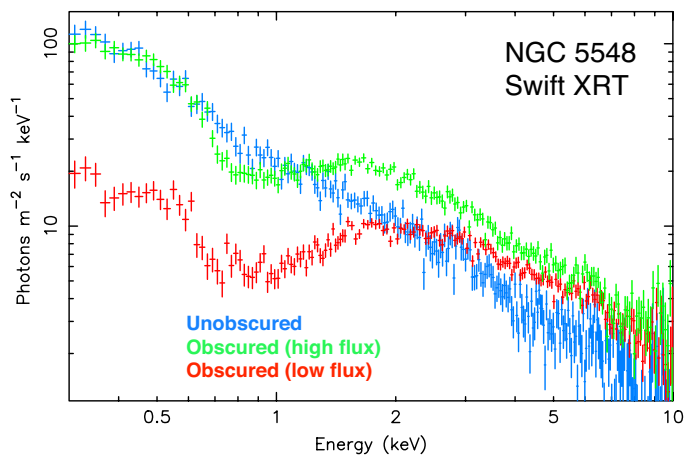}}
\vspace{-1.0cm}
\caption{Examples of \ngc\ \swift XRT spectra taken at different epochs. The unobscured spectrum (shown in blue) corresponds to the archival 2005 and 2007 data. The low-flux obscured spectrum (shown in red) is produced from observations taken between 22 June and 1 August 2013 (56465--56505 in MJD). The high-flux obscured spectrum (shown in green) corresponds to observations taken between 20 March and 29 April 2014 (56736--56776 in MJD). The visible spectral changes are indicative of variability induced by both the obscurer and the underlying continuum emission.}
\label{xrt_spectra}
\end{figure}

Similar to the lightcurves, an overview of the \swift X-ray spectra also exhibits the effects of obscuration and its variability. Figure \ref{xrt_spectra} shows examples of XRT spectra taken at different epochs highlighting interesting spectral changes. The observed spectral variability can be due to a combination of the continuum and the obscuration variability, which produces a characteristic dip and curvature in the soft X-ray band. Compared to the unobscured spectrum, which still includes absorption by a traditional warm absorber \citep{Ste05}, the dominant absorption effects by the obscurer are obvious in the 2013 and 2014 obscured spectra. However, in the low-flux obscured spectrum (which overlaps with our summer 2013 \xmm campaign), the absorption by the obscurer appears stronger than in later observations taken in spring 2014. In Sect. \ref{broadband_sect}, we investigate the origin of these observed features in the \swift lightcurves and spectra of \ngc by modelling the variability of the obscurer and the underlying broadband continuum.

The exposure time of a \swift observation is short (1--2 ks), so in order to improve the signal-to-noise of the XRT spectra and hence improve constraints on the parameters of our spectral model, time-averaged stacked spectra were created. By examining the \swift data of \ngc, we found that 10 days is the optimum interval for stacking, providing XRT spectra with both sufficient statistics and near continuous equally time-binned datasets. Therefore, in our broadband spectral modelling of Sect. \ref{broadband_sect}, the \swift XRT and UVOT data, as well as the HST COS continuum data, were time-averaged over 10 days. Similarly, in our temporal rms variability analysis of Sect. \ref{RMS_sect}, a time sampling bin size of 10 days was selected.

%

\section{Broadband spectral modelling of the {\textbf{\emph{Swift}}} data}
\label{broadband_sect}

We modelled the \swift data of \ngc using the broadband continuum model derived in \citetalias{Meh14a}, which was obtained from modelling of the stacked simultaneous data (\xmm, \nustar, \integral, HST COS and two ground-based optical observatories) from our campaign of summer 2013. In this global model of the SED, which covers energies from near-infrared (NIR) (wavelength of 8060\ \AA) to hard X-rays (200\ keV), we take into account various non-intrinsic emission and absorption processes along our line of sight, which are: Galactic reddening, host galaxy stellar emission, BLR and narrow-line region (NLR) emission lines, blended \ion{Fe}{ii} and Balmer continuum, Galactic interstellar X-ray absorption, the traditional warm absorber, soft X-ray emission lines and finally absorption by the obscurer (see \citetalias{Meh14a} for more details). The parameters of these model components, except those of the obscurer, have been kept fixed throughout our modelling of the \swift data.

The broadband continuum of our model consists of a Comptonisation component for the NIR-optical-UV continuum and the soft X-ray excess, and a power-law component for the primary hard X-ray continuum, plus a neutral X-ray reflection component. As justified in \citetalias{Meh14a}, warm Comptonisation is a feasible and likely explanation for the soft X-ray excess emission in \ngc. This model can also explain the apparent correlation between the soft X-ray and UV flux (Sect. \ref{swift_lc_sect}). In this model the thermal seed photons from the disk are up-scattered in a warm, optically thick corona to produces the soft excess emission (see e.g. \citealt{Done12}). So the NIR-optical-UV continuum of \ngc is composed of a single thermal Comptonised component ({\tt comt} in {\tt SPEX}), which also produces the soft X-ray excess (\citetalias{Meh14a}). The up-scattering Comptonising plasma was set to have a disk geometry and the initial values for its parameters were fixed to those obtained from the stacked 2013 spectra in \citetalias{Meh14a}: temperature of the seed photons $T_{\rm seed} = 0.80$~eV, electron temperature of the plasma $T_{\rm e} = 0.17$~keV, optical depth $\tau = 21.1$ with {\tt comt} normalisation of $6.0 \times 10^{55}$ photons $\mathrm{s}^{-1}\ \mathrm{keV}^{-1}$.

In addition to the Comptonised soft excess component, the primary hard X-ray continuum was modelled with a cut-off power-law ({\tt pow} in {\tt SPEX}), which mimics the spectral emission produced by inverse Compton scattering in an optically thin (${\tau \sim 1}$), hot (${T_{\rm{e}} \sim 100\ {\rm keV}})$ corona (e.g. \citealt{Suny80,Haar93}). The cut-off power-law is known to provide a reasonable approximation for the hot Comptonisation spectrum (e.g. \citealt{Petr00}). The weak reflection component in \ngc was also included in our modelling, which is consistent with being constant and produced in neutral, distant material \citep{Ursin14,Cappi14}. Since we can only fit \swift XRT data up to 10~keV, all the parameters of the reflection spectrum and the high-energy exponential cut-off of the power-law (400 keV), which are not in the XRT bandpass, have been fixed to those reported in \citetalias{Meh14a}. For the reflection component, we used the {\tt refl} model in {\tt SPEX}, which computes the reflected continuum and the corresponding \FeKa line from a constant density X-ray illuminated atmosphere. At low energies, the power-law was smoothly broken before overshooting the energies of the seed photons from the disk as described in \citetalias{Meh14a}.

The obscurer model, first described in \citetalias{Kaas14}, consists of two photoionised absorption components, each modelled with an {\tt xabs} component in {\tt SPEX}. The {\tt xabs} model calculates the transmission through a slab of photoionised gas, where all ionic column densities are linked in a physically consistent fashion through the {\tt Cloudy} \citep{Fer98} photoionisation model. The parameters of an {\tt xabs} component are the ionisation parameter ($\xi$), the equivalent hydrogen column density ($\NH$), the covering fraction $C_f$ of the absorber, its flow $v$ and turbulent $\sigma_v$ velocities. The ionisation parameter $\xi$ \citep{Tar69} is defined as ${\xi \equiv {L}\, /\, {{n_{\rm{H}} r^2 }}}$, where $L$ is the luminosity of the ionising source over the 1--1000 Ryd (13.6 eV to 13.6 keV) band in $\rm{erg}\ \rm{s}^{-1}$, $n_{\rm{H}}$ the hydrogen density in $\rm{cm}^{-3}$ and $r$ the distance between the ionised gas and the ionising source (in cm). In this time-averaged model of the obscurer, component 1 covers about 86\% of the central X-ray emitting region, with $\log \xi = -1.2$ and ${\NH = 1.2 \times 10^{22}\ \rm{cm}^{-2}}$. Component 2 of the obscurer covers 30\% of the X-ray source and is almost neutral ($\log \xi = -4.0$) with ${\NH = 9.6 \times 10^{22}\ \rm{cm}^{-2}}$. We shall refer to component 1 and 2 as warm and cold phases of the obscurer, respectively. For the unobscured epoch (2005 and 2007 \swift data), the obscurer components were excluded from our model by setting $C_f$ of both obscurer components to zero. During the obscured epoch (February 2012 onwards) we set the initial parameters of the obscurer to those of \citetalias{Kaas14} and \citetalias{Meh14a}. Since the obscurer is located between the central ionising source and the traditional warm absorber, it prevents some of the ionising EUV and X-ray radiation from reaching the warm absorber. Thus the different phases of the warm absorber become less ionised (de-ionised) and induce more X-ray absorption than when \ngc was unobscured. In our modelling of the obscured \swift data we used the de-ionised warm absorber model obtained by \citetalias{Kaas14}, which consists of six phases ({\tt xabs}) of photoionisation. For the unobscured \swift data from 2005 and 2007, the warm absorber would have been exposed to a normal unobscured ionising SED, whose parameters are derived by \citetalias{Kaas14} and used in our modelling here.

%

%
\begin{figure}[!tbp]
\centering
\vspace{-0.8cm}
\resizebox{1.1\hsize}{!}{\hspace{-0.8cm}\includegraphics[angle=0]{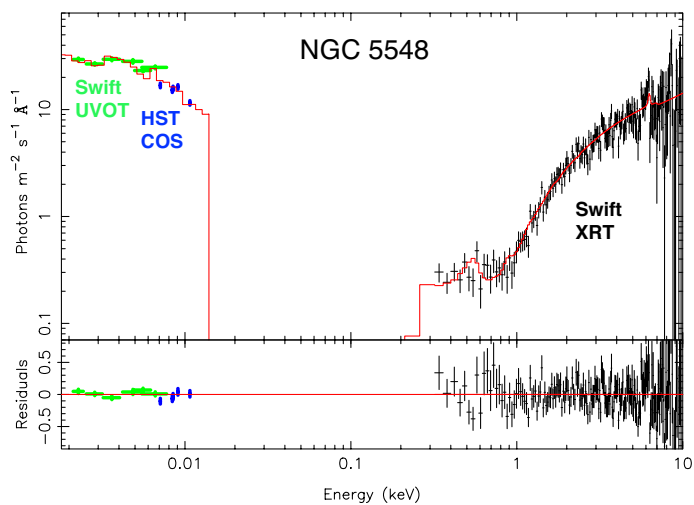}}\vspace{-0.3cm}
\caption{A typical example of one set of observed optical/UV (\swift UVOT and HST COS) and X-ray (\swift XRT) spectra of \ngc, which were fitted with the broadband spectral model described in Sect. \ref{broadband_sect}. The UVOT fluxes are from six photometric filters and the HST COS fluxes are from five narrow energy bands in the UV continuum. Residuals of the fit, defined as (observed$-$model)/model, are shown in the bottom panel. The displayed data are from a 10-day bin period between 2--12 March 2014.}
\label{swift_fit}
\end{figure}

We started our spectral fitting procedure by first fitting the 3.0--10 keV part (which is almost unabsorbed) of each of the 93 XRT spectra by freeing the $\Gamma$ and normalisation of the power-law component ({\tt pow}). The power-law parameters were then temporarily kept frozen while the whole 0.3--10 keV XRT spectrum, and optical/UV (UVOT and HST COS) data were fitted by freeing the normalisation of the Comptonisation ({\tt comt}) component. This was then followed by freeing the covering fraction of the warm phase of the obscurer (\cf), which from the analysis of \xmm spectra by \citet{DiGesu14b} and \citet{Ursin14} is found to be significantly variable. Then, the power-law parameters were freed again to obtain a best-fit.

Next, we tested whether fitting other parameters of the obscurer improves our spectral fits. By freeing the covering fraction of the cold phase of the obscurer (which has a much lower value than that of the warm phase), our fits did not improve and the covering fraction often became unconstrained. We thus kept the covering fraction of the cold phase of the obscurer fixed to its initial value throughout our modelling as XRT spectrum is much less sensitive to this weaker phase than the more prominent warm phase of the obscurer. We discuss the significance of this assumption in Sect. \ref{discuss_obsc}. Hereafter, obscurer \cf refers to the covering fraction of the warm phase of the obscurer. We then checked whether freeing the column densities \NH of the obscurer components improves our fits. However, we did not observe any significant improvement in our fits and the \NH values became weakly constrained. We then tried another scenario, in which the covering fractions of the components were fixed to their initial values, and instead their \NH were fitted. But in this case, for most datasets we obtained a worse fit. So in our modelling we have kept \NH of the obscurer components fixed to the best-fit values obtained by \citetalias{Kaas14} from the stacked summer 2013 spectra, and allowed \cf of the obscurer to vary. In Sect. \ref{RMS_sect} we will show that the rms variability produced by changes in the \cf of the obscurer is indeed consistent with the observed rms spectra of \ngc.

Next, we checked for possible variability of $T_{\rm seed}$, $T_{\rm e}$ and $\tau$ of the {\tt comt} component, and whether freeing these parameters improve our broadband spectral fits. We found that freeing $T_{\rm seed}$ helps in making a better fit to the optical/UV (UVOT and HST COS) data in some observations, particularly when the UV flux had a flare-up. We thus left $T_{\rm seed}$ as a free parameter in our modelling. On the other hand, freeing $T_{\rm e}$ and $\tau$ did not make a statistically significant difference to our fits. Therefore, we froze $T_{\rm e}$ and $\tau$ to avoid unnecessary free parameters and kept them fixed to their initial values. Indeed varying the normalisation and $T_{\rm seed}$ of {\tt comt} is sufficient to provide a good fit to the optical/UV data and the soft X-ray excess in all our datasets. At the final step, the goodness of our fits was re-examined. In all the observations, the data were well fitted with good C-statistic values. In addition to examining the C-statistic output, the quality of our best-fits was also inspected visually at the end to make sure all optical/UV and X-ray data were properly fitted.

\begin{figure*}[!tbp]
\centering
\vspace{-0.1cm}
\resizebox{0.825\hsize}{!}{\includegraphics[angle=0]{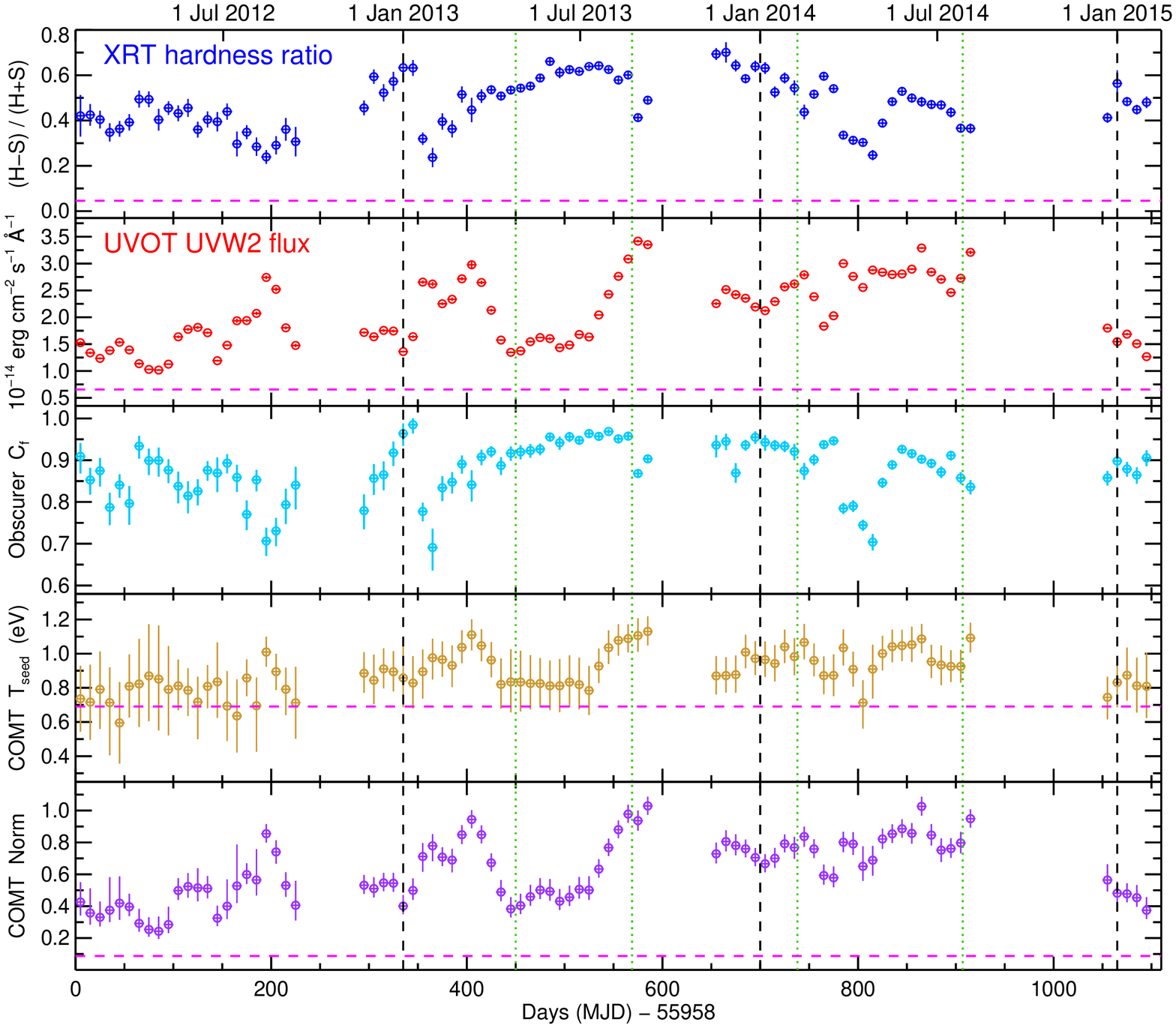}}\vspace{-0.2cm}
\resizebox{0.825\hsize}{!}{\includegraphics[angle=0]{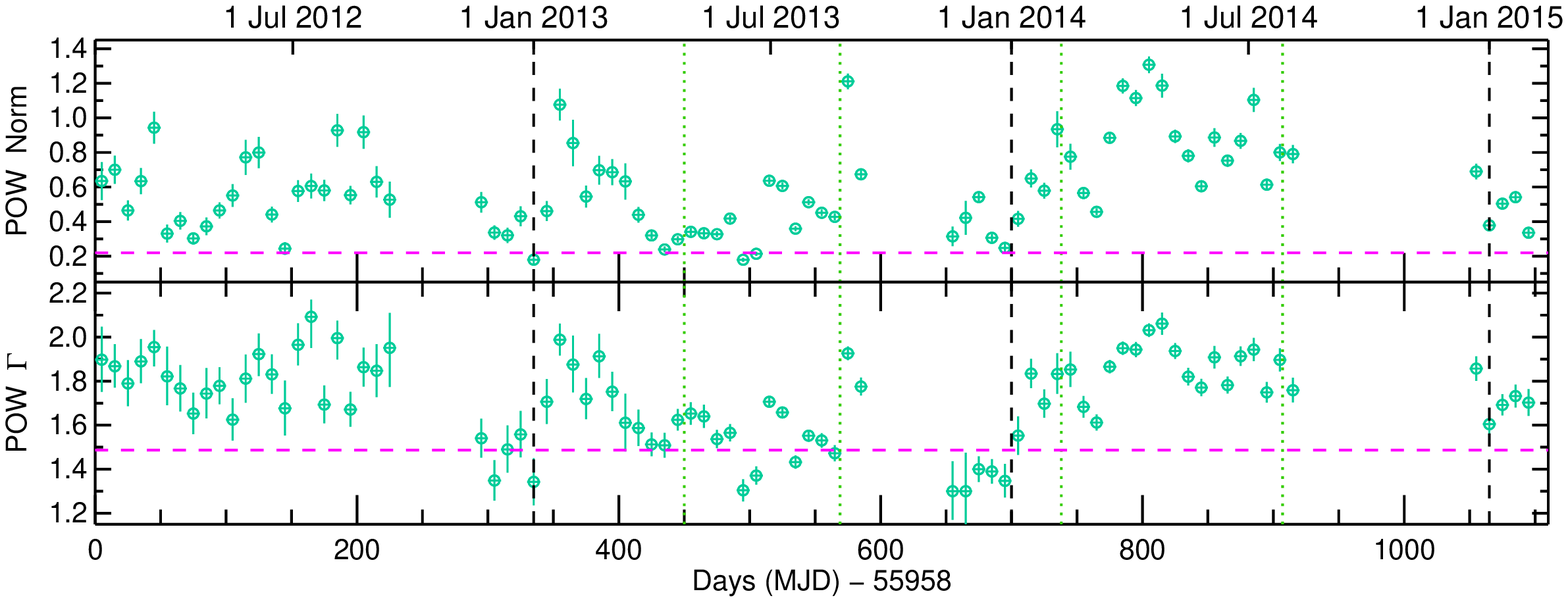}}\vspace{-0.2cm}
\resizebox{0.825\hsize}{!}{\includegraphics[angle=0]{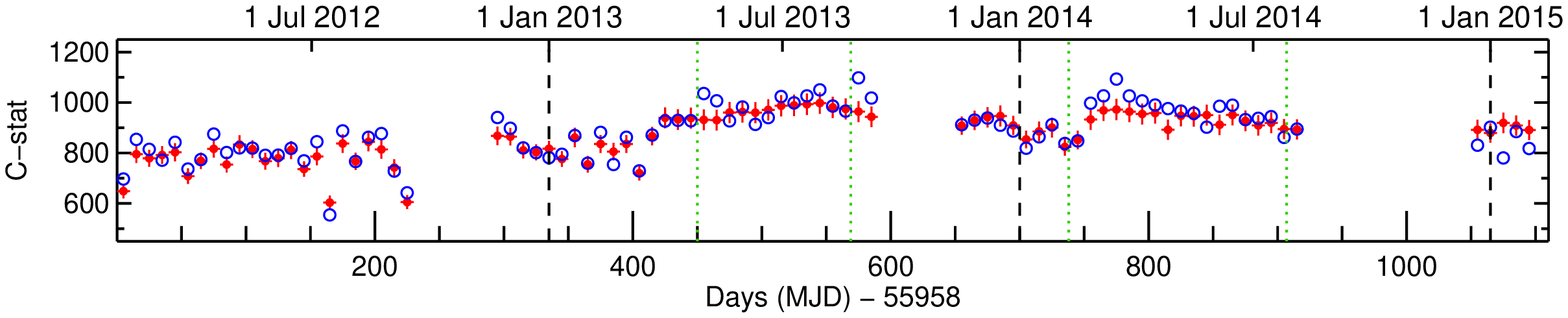}}\vspace{-0.1cm}
\caption{Variability of the parameters derived from broadband spectral modelling of \swift and HST COS data of \ngc in Sect. \ref{broadband_sect}. The X-ray hardness ratio $R$ and the UVW2 flux are also displayed on the top two panels for comparison. On the vertical axis of each panel the name of the corresponding parameter is given. The normalisation of the warm Comptonisation ({\tt comt}) component (modelling the optical/UV continuum and the soft X-ray excess) is shown in units of $10^{56}$ photons s$^{-1}$ keV$^{-1}$. The power-law ({\tt pow}) normalisation is in $10^{52}$~photons~s$^{-1}$~keV$^{-1}$ at 1 keV. In the bottom panel, the goodness of the fits are shown, with blue open circles indicating the obtained C-stat and the filled red circles indicating the expected C-stat with its rms uncertainties. The range of the above time-series is from 1 February 2012 to 4 February 2015. The vertical dotted lines in green indicate two interesting periods in 2013 and 2014, which were selected for the temporal variability analysis of Sect. \ref{RMS_sect} and production of rms spectra in Fig. \ref{rms_spectra}. The vertical dashed lines in black indicate the start of each year for reference. The horizontal dashed lines in magenta represent the averages from previous years (2005 and 2007 \swift data from the unobscured epoch) derived in this work with the same model.}
\label{par_lc}
\end{figure*}

In Fig. \ref{swift_fit}, we show a typical example of one set of \swift (XRT and UVOT) and HST COS spectral data (time-averaged over 10 days), together with its best-fit broadband model. {To find out how individual model spectral components contribute to the global broadband model used in this paper, see Figs. 5 and 10 in \citetalias{Meh14a}.} In total, a series of 93 sets of such stacked data were fitted using a batch processing script incorporating our broadband model described above. Figure \ref{par_lc} illustrates the variability of the fitted parameters of our broadband spectral model of the continuum and obscuration. The obtained and expected C-statistic values (see the {\tt SPEX} manual for their definitions) are also provided in this figure. Additionally, the X-ray hardness ratio $R$ and the observed flux in the UVOT UVW2 filter are also displayed for comparison with the derived parameters. We examine the variability of the model parameters of Fig. \ref{par_lc} in Sect. \ref{para_results}.

%

\subsection{Long-term variability of the obscuration and continuum parameters}
\label{para_results}

\begin{figure}[!tbp]
\centering
\resizebox{1.0\hsize}{!}{\hspace{-0.5cm}\includegraphics[angle=0]{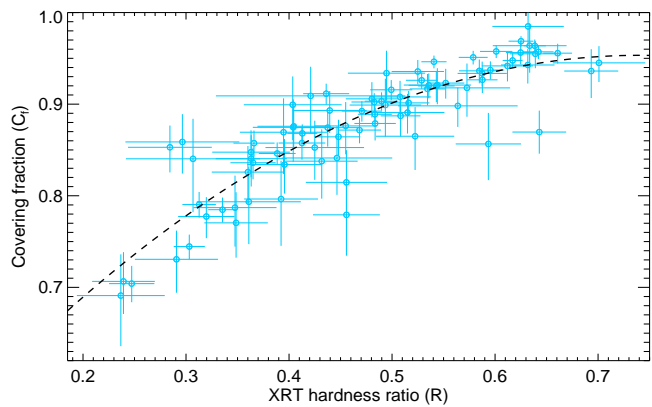}}
\caption{Relation between the covering fraction ($C_f$) of the obscurer and the observed XRT hardness ratio ($R$). The data have been fitted with Eq. \ref{cf_eq}, shown in dashed black line, as described in Sect. \ref{para_results}.}
\label{obscurer_fcov}
\end{figure}

The time-series of model parameters in Fig. \ref{par_lc} exhibit interesting features, which we examine here. The variability trend of the obscurer covering fraction \cf is very similar to that of the X-ray hardness ratio $R$ measured with \swift XRT. For correlation between \cf and $R$ in our obscured sample size of 85, the Pearson correlation coefficient is ${r = 0.858}$. This gives a negligible null hypothesis probability of ${p_{\rm null} < 10^{-10}}$. Thus, there is a strong positive correlation between \cf and $R$. In Fig. \ref{obscurer_fcov}, we have plotted the obscurer \cf versus $R$. The data are well fitted with a quadratic function, given by:

\begin{equation}
\label{cf_eq}
{C_{f} =  0.46 + 1.34\ R - 0.91\ R^{2}} {\rm \ \ \ \ \ \ \ for \ \ \ } 0.7 \le {C_{f}} \le 1
\end{equation}
We note that a quadratic function is preferred over a linear one as it provides a better description of the data at both low and high extremes.

%

Since the start of the \swift monitoring in February 2012, the \cf time-series in Fig. \ref{par_lc} shows that at no point in time the central X-ray source became unobscured. The obscuration has been continuously present, with its covering fraction varying over time between 0.70 and 0.98. A useful aspect of the relation in Fig. \ref{obscurer_fcov} is that by simply measuring the hardness ratio $R$ from \swift observations, one roughly knows the covering fraction of the obscurer without any modelling. We note that apart from the clear relation between \cf and $R$, there is also a weaker anti-correlation (${p_{\rm null} \approx 0.05}$) between these parameters and the UVW2 flux when it flares up (steep continuous increase over 1--2 months). The increase in the UV flux is accompanied by a decrease in the covering fraction of the obscurer. We note that this anti-correlation contributes in producing the scatter observed between the \cf and $R$ relation in Fig. \ref{obscurer_fcov}. We further discuss these correlations and the obscurer \cf variability in Sect. \ref{discussion}.

\begin{figure}[!tbp]
\centering
\resizebox{1.0\hsize}{!}{\hspace{-0.5cm}\includegraphics[angle=0]{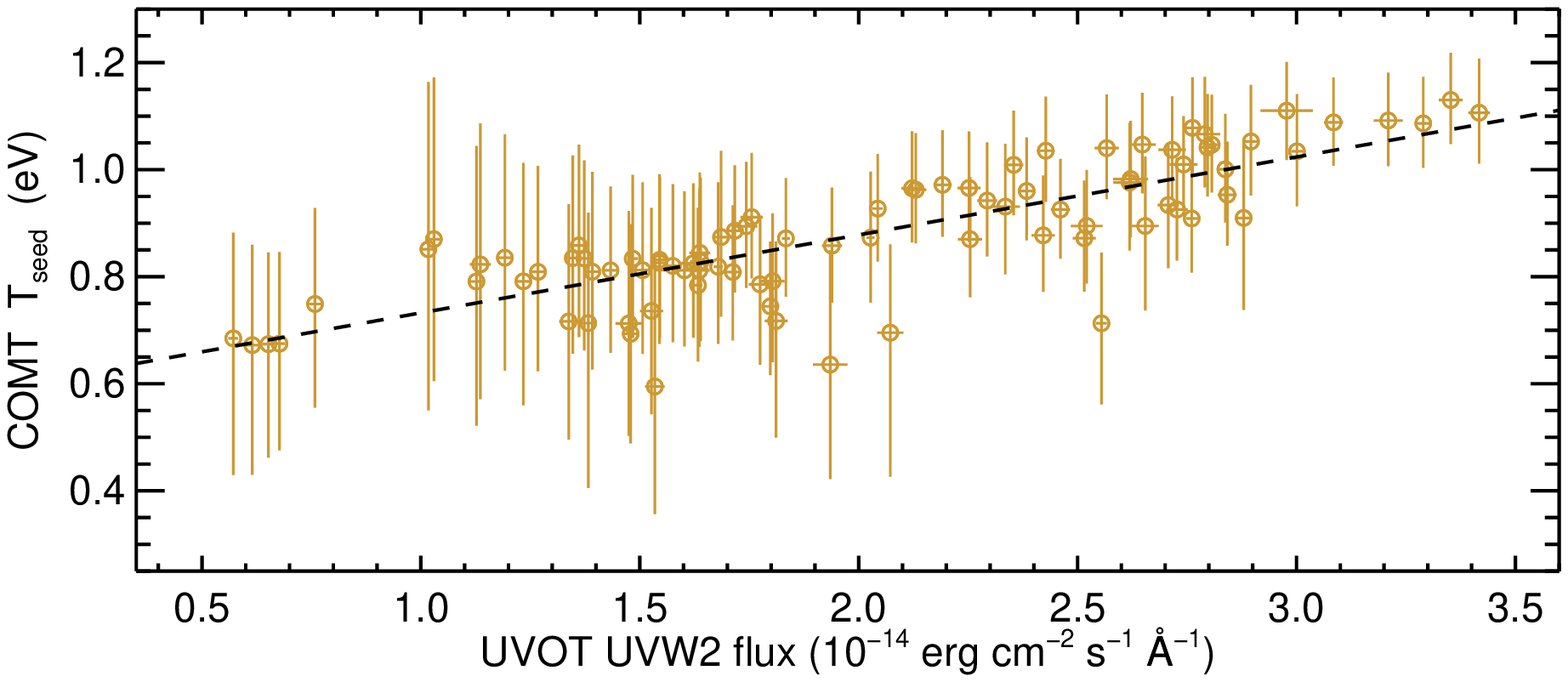}}\vspace{-0.4cm}
\resizebox{1.0\hsize}{!}{\hspace{-0.5cm}\includegraphics[angle=0]{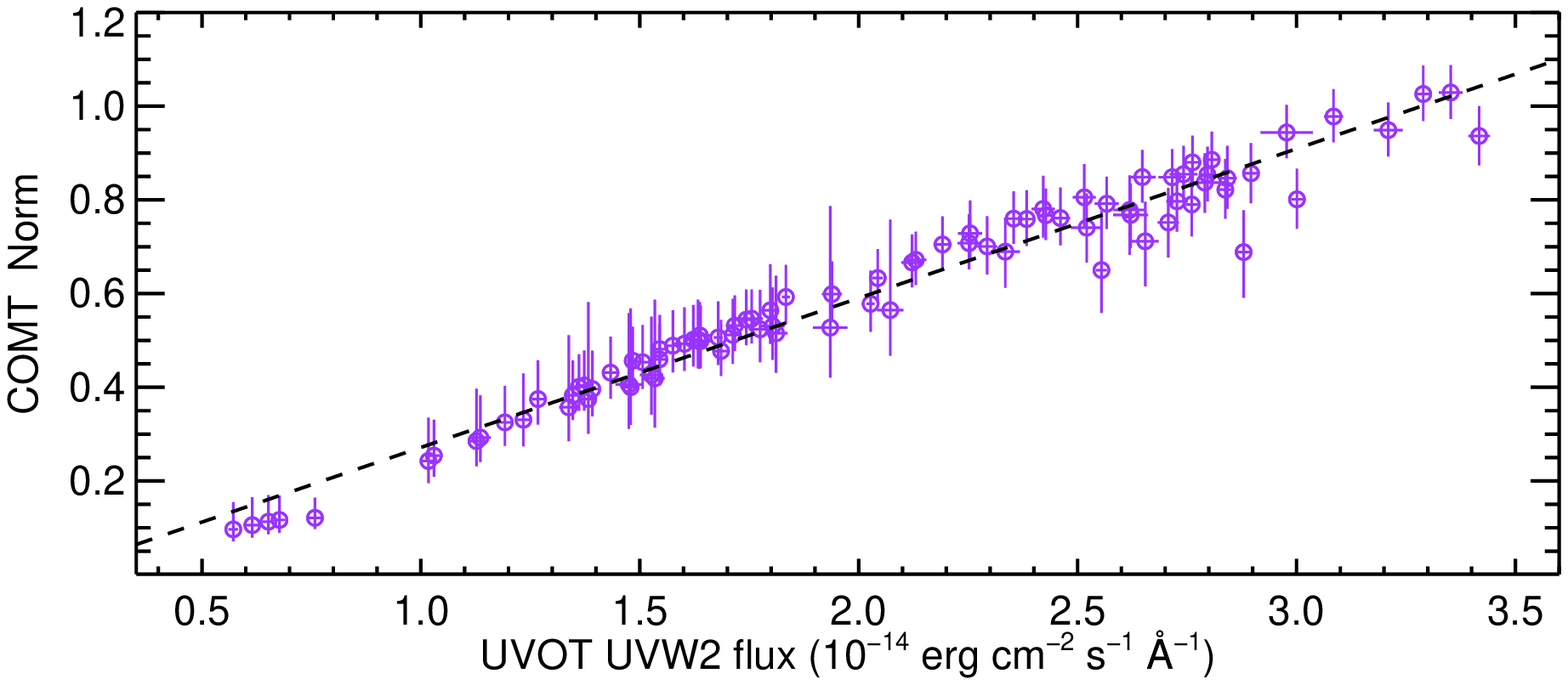}}
\caption{{\it Top panel}: Relation between the seed photon temperature $T_{\rm seed}$ of the {\tt comt} (Comptonised optical-UV/soft X-ray excess) component and the observed UV flux in the UVW2 filter. {\it Bottom panel}: Relation between the normalisation of {\tt comt} and the observed UVW2 flux. The {\tt comt} normalisation is shown in units of $10^{56}$ photons~s$^{-1}$ keV$^{-1}$. The data in the above panels have been fitted with linear functions, shown in dashed black lines, as described in Sect. \ref{para_results}.}
\label{COMT_uv_rel}
\end{figure}

The time-series of Fig. \ref{par_lc} also show a clear link between the variability of the UVW2 flux and the parameters of the warm Comptonisation component ({\tt comt}). As this is the component which fits the optical/UV continuum and the soft X-ray excess, there is naturally a tight correlation between the normalisation of {\tt comt} ($N_{\tt comt}$) and the UV flux. However, there also appears to be a correlation between the UVW2 flux and $T_{\rm seed}$ of {\tt comt} when the UVW2 flux flares up. However, at other times, $T_{\rm seed}$ is more or less unchanged within its larger error bars as $N_{\tt comt}$ alone is sufficient to fit the optical/UV variability in most observations. In Fig. \ref{COMT_uv_rel}, the relation between the UVW2 flux and $T_{\rm seed}$ (top panel), and normalisation of {\tt comt} (bottom panel) are shown. For a sample size of 90 (which have UVW2 data in our spectral modelling), the correlation coefficient between $F_{\rm{UVW2}}$ and $T_{\rm seed}$ is $r = 0.810$, and between $F_{\rm{UVW2}}$ and ${N_{\tt comt}}$ is $r = 0.981$, with both giving a negligible $p_{\rm null}$ (${< 10^{-10}}$), indicating strong positive correlations. These relations are described well with linear functions: ${{T_{\rm seed}} = 0.59 + 0.15\ F_{\rm{UVW2}}}$ and ${{N_{\tt comt}} = - 0.047 + 0.32\ F_{\rm{UVW2}}}$, where the unit of ${T_{\rm seed}}$ is in eV, $N_{\tt comt}$ in $10^{56}$~photons~s$^{-1}$~keV$^{-1}$ and the observed $F_{\rm{UVW2}}$ in ${10^{-14}\ \ergcm}$.

%

%
\begin{figure}[t]
\centering
\resizebox{1.0\hsize}{!}{\hspace{-0.5cm}\includegraphics[angle=0]{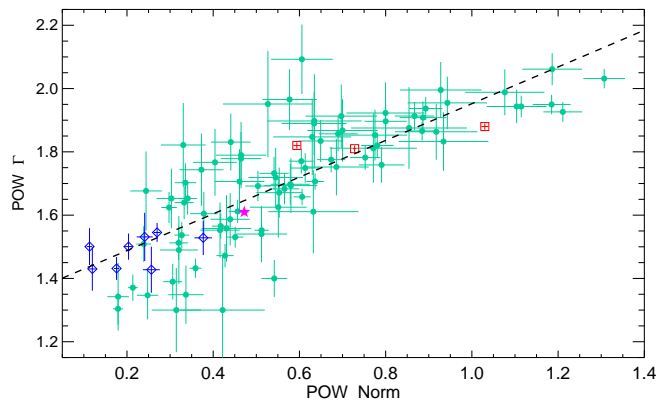}}
\caption{Relation between the photon index $\Gamma$ and the normalisation of the power-law ({\tt pow}) continuum as described in Sect. \ref{para_results}. The {\tt pow} normalisation is shown in units of $10^{52}$~photons~s$^{-1}$ keV$^{-1}$ at 1 keV. The data in green filled circles represent the \swift data from the obscured epoch (Feb 2012 onwards), the blue open diamonds are the \swift data from the unobscured epoch of 2005 and 2007, the red open squares represent three \xmm observations from unobscured epoch of 2000 and 2001, and the single filled magenta star corresponds to the stacked data from our summer 2013 campaign when the source was obscured. The data have been fitted with Eq. \ref{pow_rel_eq} shown in dashed black line.}
\label{pow_gam_norm}
\end{figure}

The parameter variability results of Fig. \ref{par_lc} also show that there is a relation between $\Gamma$ and the normalisation of the power-law ($N_{\tt pow}$), with the power-law spectrum becoming softer as it gets brighter. In Fig. \ref{pow_gam_norm} we have plotted $\Gamma$ versus the normalisation of the power-law. In this figure, as well as the {\tt pow} parameters derived from our modelling of the \swift data, we have also shown the {\tt pow} parameters from 2000 and 2001 \xmm observations and from the stacked summer 2013 data of our campaign, obtained in \citetalias{Meh14a}. The correlation coefficient between $N_{\tt pow}$ and $\Gamma$ is $r = 0.801$ for a sample size of 97, which gives a negligible $p_{\rm null}$ (${< 10^{-10}}$), indicating a strong positive correlation. The data shown in Fig. \ref{pow_gam_norm} are fitted with the linear function:
\begin{equation}
\label{pow_rel_eq}
{\Gamma = 1.37 + 0.58\ N_{\tt pow}}
\end{equation}
where normalisation $N_{\tt pow}$ is in units of $10^{52}$~photons~$\mathrm{s}^{-1}$ $\mathrm{keV}^{-1}$ at 1 keV.

Finally, we note that there is a significant correlation between the UVW2 flux ($F_{\rm{UVW2}}$) and the power-law parameters. For the sample size of 90, the correlation coefficient between $F_{\rm{UVW2}}$ and $\Gamma$ is ${r = 0.288}$, and between $F_{\rm{UVW2}}$ and $N_{\tt pow}$ is ${r = 0.645}$. The corresponding null hypothesis probability for these correlations is ${p_{\rm null} = 0.0059}$ and ${p_{\rm null} < 10^{-10}}$, respectively, indicating positive correlations between the variability of the UV flux and the power-law component.

\section{Broadband temporal rms variability analysis}
\label{RMS_sect}

In this section we use the \swift and HST COS lightcurves of \ngc to determine the rms variability at different energies. The shape of an rms spectrum provides us with an important model-independent insight into the nature of variability. We can then compare the observed rms spectra of \ngc with the model rms spectra generated from the results of our broadband spectral modelling in Sect. \ref{broadband_sect} (Fig. \ref{par_lc}).

We calculated the fractional rms variability amplitude, following the recipe of \citet{Vaugh03}. For a series of $N$ flux measurements of ${x_i}$, with a mean of ${\overline x}$, the fractional rms variability amplitude \fvar is given by
\begin{equation}
\label{fvar_eq}
{F_{{\mathop{\rm var}} }} = \sqrt {\frac{{\sigma _{{\rm{XS}}}^2}}{{{{\overline{x}\,}^2}}}}
\end{equation}
where the excess variance ${\sigma^{2}_{\rm{XS}} = {{S^2} - \overline {\sigma _{{\rm{err}}}^2} }}$, the variance ${{S^2} = \sum\limits_{i = 1}^N {{{({x_i} - \overline{x}\,)}^2}} /(N - 1)}$ and ${\overline {\sigma _{{\rm{err}}}^2} }$ is the mean square error on the count rate measurements. To compute the uncertainty of \fvar, we used the prescription developed in \citet{Vaugh03}, which utilises Monte Carlo simulations to take into account the effects of flux measurement errors; see Eq. 11 and Appendix B in \citet{Vaugh03} for more details. Time sampling bin size ${\Delta t_{\rm{bin}}}$ of 10 days was selected for our computations to match the $\Delta t_{\rm{bin}}$ of our broadband spectral modelling in Sect. \ref{broadband_sect}. 

We calculated \fvar for the \ngc lightcurves at 25 different energy bins: 14 \swift XRT energy bins between 0.3--10 keV, the 6 UVOT photometric filters (V, B, U, UVW1, UVM2 and UVW2) and the 5 HST COS continuum measurements from narrow wavelength bins in the UV (1158 \AA, 1367 \AA, 1462 \AA, 1479 \AA\ and 1746 \AA). The rms variability spectra were calculated for lightcurves from two particular periods in 2013 and 2014: 26 April 2013 to 23 August 2013 and 8 February 2014 to 27 July 2014. These periods, which are indicated by the green vertical dotted lines in the panels of Fig. \ref{par_lc}, were selected for two main reasons: (1) they correspond to the most intense and continuous \swift monitorings of \ngc, overlapping with the \xmm campaign of 2013 and the HST COS monitoring program of 2014 (thus also making the construction of COS rms variability spectra possible); (2) the results of our broadband spectral modelling in Fig. \ref{par_lc} suggest different variability behaviour for key parameters during these two epochs, which are interesting to explore: the UV flux and the {\tt comt} component display a more steady and larger increase over time in 2013 than in 2014; on the other hand, the obscurer \cf varies over a larger range in 2014 than in 2013. Thus, examining the differences in the shape of the observed rms variability spectra over these periods provides a useful model-independent signature for the nature of the variability, and we can check whether the observed rms spectra are consistent with our broadband modelling results.

%
\begin{figure*}[!tbp]
\centering
\resizebox{0.7\hsize}{!}{\hspace{-0.7cm}\includegraphics[angle=0]{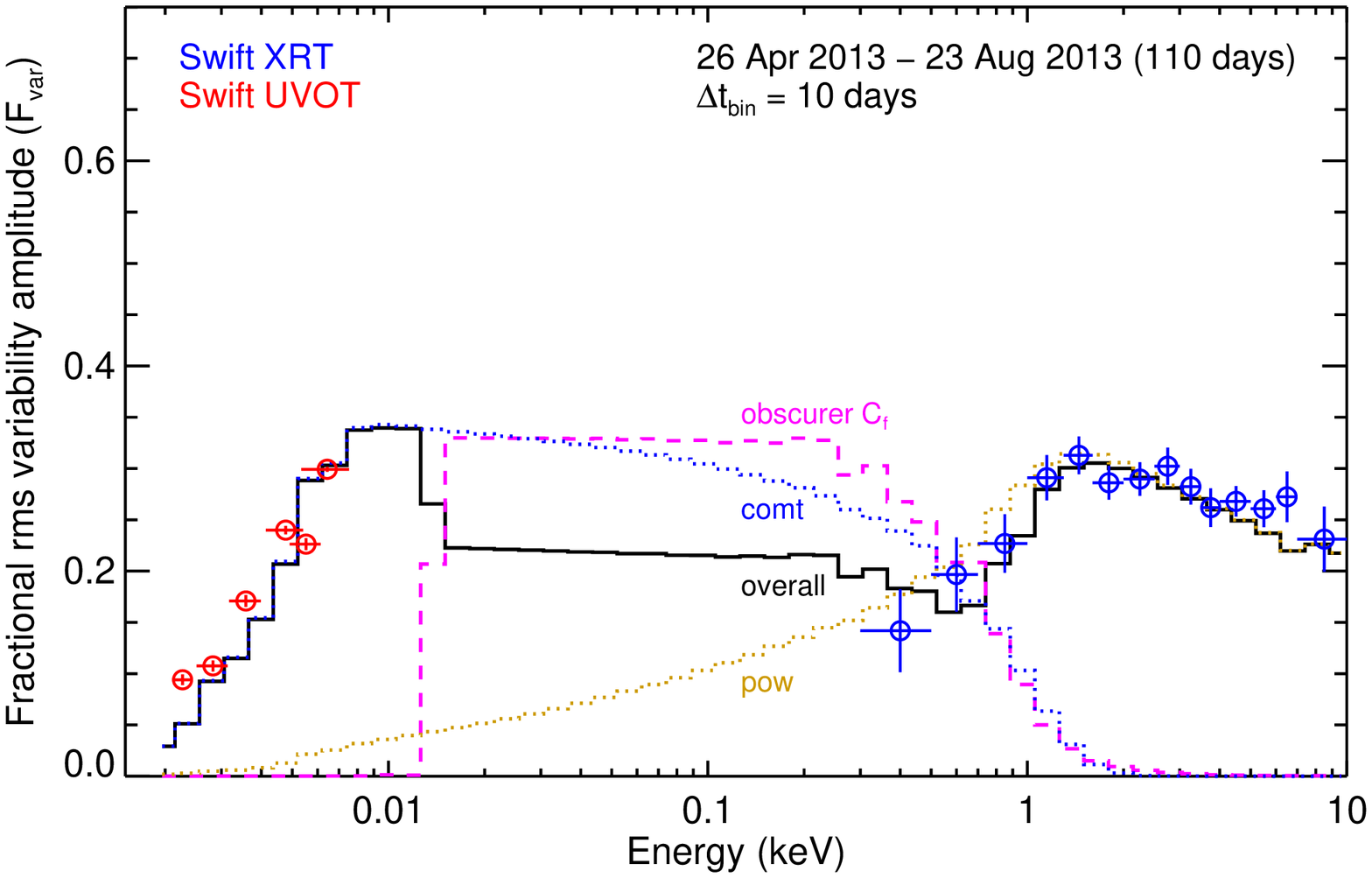}}\vspace{-0.5cm}
\resizebox{0.7\hsize}{!}{\hspace{-0.7cm}\includegraphics[angle=0]{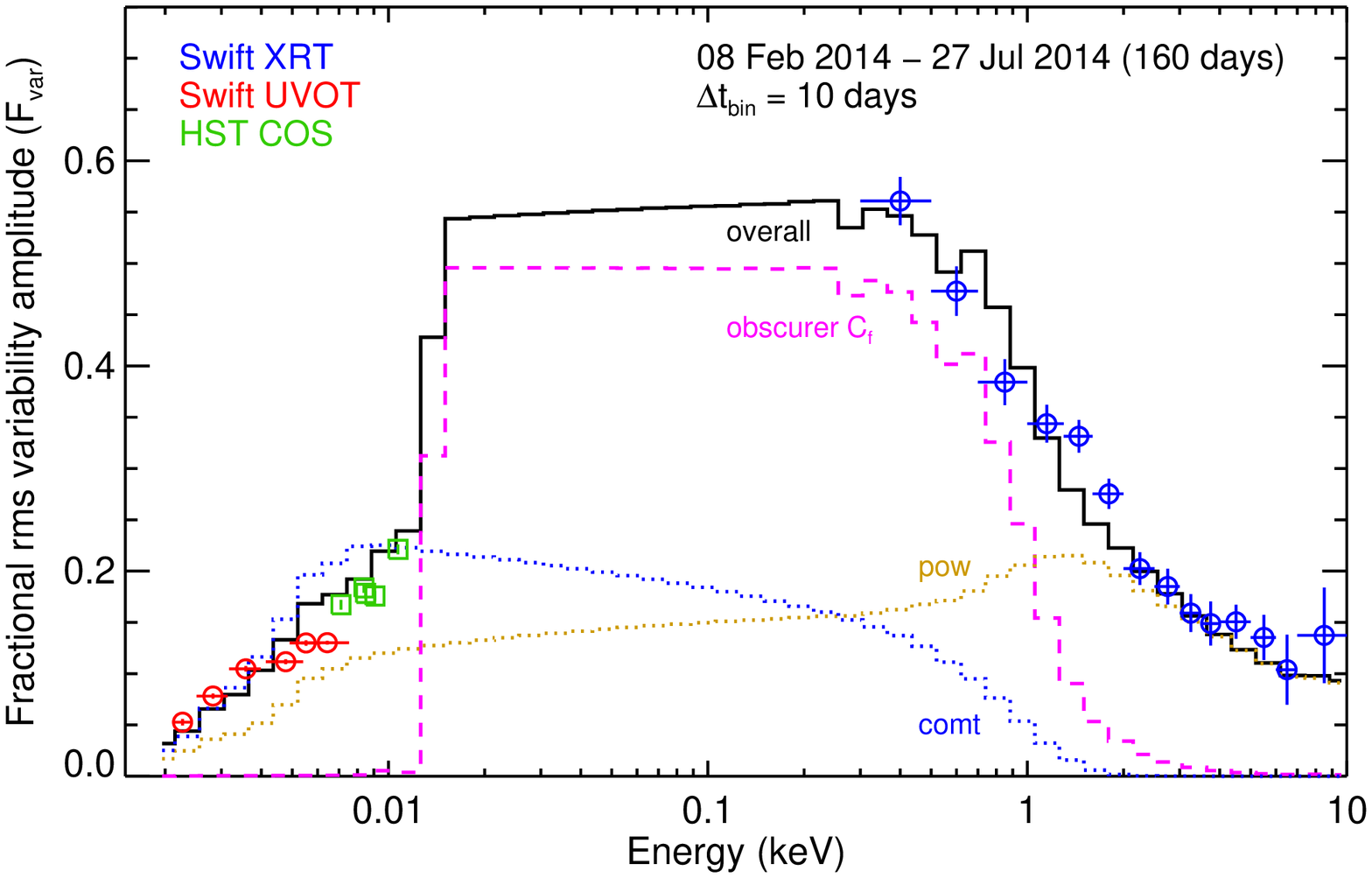}}
\caption{\swift and HST COS broadband rms variability spectra of \ngc for selected periods in 2013 ({\it top panel}) and 2014 ({\it bottom panel}). The XRT data are shown in blue circles and the UVOT data in red circles. The HST COS data (available only for the 2014 period) are displayed in green squares in the bottom panel. The overall rms variability model generated from the variability of all the parameters derived from our spectral modelling (Sect. \ref{broadband_sect} and Fig. \ref{par_lc}) is over-plotted in solid black line on the data. The rms variability of individual model components are also over-plotted in the panels for comparison: {\tt comt} in blue dotted line, {\tt pow} in brown dotted line and the obscurer \cf in dashed magenta line. The calculation of the rms spectra are described in Sect. \ref{RMS_sect}. The axes ranges in both panels are identical.}
\label{rms_spectra}
\end{figure*}

Figure \ref{rms_spectra} shows the rms variability spectra of \ngc for the selected periods in 2013 and 2014. In addition to the observed rms spectra from \swift and HST COS over the selected periods, we have over-plotted the model rms spectra generated from the results of our broadband spectral modelling in Sect. \ref{broadband_sect} (Fig. \ref{par_lc}). The overall model rms spectrum (shown in black solid line in each panel) is calculated by letting all the free parameters of the model to vary within the observed best-fit values shown in Fig. \ref{par_lc} (i.e. the parameters of {\tt pow}, {\tt comt}, as well as the obscurer \cf). The overall model rms spectrum agrees well with the observed rms spectrum in both 2013 and 2014 periods, confirming that the results of our broadband spectral modelling in Sect. \ref{broadband_sect} matches our model-independent results from the rms variability analysis of the count rate lightcurves. Apart from the overall model rms spectra shown in Fig. \ref{rms_spectra}, we have also derived the rms spectrum of individual components of our broadband spectral model ({\tt comt}, {\tt pow} and the obscurer \cf), which help us in visualising the spectral variability of these components. The rms spectrum for each component was calculated by fixing all the parameters of the model to that of the first observation in each period and then letting only the interested parameters (e.g. \cf) to vary according to the best-fit values obtained in Fig. \ref{par_lc}. We note that the overall rms spectrum is not simply the sum of individual rms spectra of the components. Depending on how the individual parameters vary over time, their combined effect on the resulting rms spectrum can be different. The rms variability spectra of {\tt comt}, {\tt pow} and the obscurer \cf are displayed in Fig. \ref{rms_spectra}.

The first striking difference between the observed rms spectra of 2013 (top panel) and 2014 (bottom panel) in Fig. \ref{rms_spectra} is the rms shape in the soft X-ray band between 0.3 and 1.0 keV. While in 2013 \fvar is low in the soft X-ray band and displays a decline towards lower energies, during the 2014 period \fvar is much higher with a steep rise towards lower energies. The second interesting difference between the two rms spectra is at optical-UV energies. While in both periods, \fvar clearly displays a linear rise extending from optical to UV energies, the rise is steeper in the UV data of 2013 than in 2014. However, beyond the Lyman limit (13.6 eV), the model rms spectrum in 2014 rises very strongly towards EUV, similar to the rise from soft X-rays towards EUV energies. 

The difference between the 2013 and 2014 observed rms spectra is due to the difference between the combined effect of the obscurer and continuum variability during these two periods. As evident in Fig. \ref{rms_spectra}, the rms variability of the obscurer \cf is dominant over those of the continuum components in the EUV and soft X-ray energies in the 2014 period, whereas it is comparable to them in the 2013 period. Note that even small variations of \cf (such as in the 2013 period) still induce significant soft X-ray flux variability. In the 2013 period, the variability of the obscurer \cf (ranging from 0.92 to 0.97 in Fig. \ref{par_lc}) and the comparable variability of the continuum components are such that their combined effect lowers the \fvar amplitude of the overall rms spectrum. On the other hand, in the 2014 period, the stronger variability of the obscurer \cf (ranging from 0.70 to 0.95 in Fig. \ref{par_lc}) and the relatively weaker variability of the continuum components are such that they result in significant increase in the \fvar amplitude of the overall rms spectrum in the EUV and soft X-ray energies. Depending on how the parameters vary over time, and any positive or negative correlations among them, their combined effect can result in a different overall rms spectrum. In both the 2013 and 2014 periods, the overall model rms spectrum agrees well with the observed rms spectrum as seen in Fig. \ref{rms_spectra}.

\section{Discussion}
\label{discussion}
%
\subsection{Variability of the X-ray obscuring outflow in \ngc}
\label{discuss_obsc}

In this investigation with \swift we have determined that the X-ray obscuring outflow in \ngc is responsible for significant variability in the soft X-ray band on long-timescales. The results of our broadband spectral modelling (Sect. \ref{broadband_sect}, Fig. \ref{par_lc}) and temporal rms variability analysis (Sect. \ref{RMS_sect}, Fig. \ref{rms_spectra}) show that this soft X-ray variability on timescales of 10 days to $\sim 5$ months can be explained by changes in the covering fraction \cf of the central X-ray source by the obscurer, which is located along our line of sight. We find that \cf has varied between 0.7 and nearly 1.0 over the last few years, since the time \swift monitoring of \ngc in the obscured state began in February 2012. The rms variability of the obscurer is found to be generally stronger than those of the power-law and the warm Comptonisation component over the EUV and soft X-ray energies. 

The \swift data point to the \cf of the obscurer being responsible for most of its long term variability. However, from analysis of the EPIC-pn data, \cite{Cappi14} find that both the covering fraction \cf and column density \NH of the obscurer components can vary on shorter timescales of days to weeks. They find the rms variability produced by \cf of the warm component of the obscurer is stronger than the variability of other parameters of the obscurer. This is the same \cf parameter, which its variability is constrained with the \swift data on long-timescales in this paper. The rms variability of other parameters of the obscurer are too subtle for detection and determination with \swift XRT due to its lower spectral resolution and effective area compared to EPIC-pn. We note that the \NH variability produces a bump with a characteristically sharp edge at 1 keV in the rms spectrum. However, this feature is not apparent in the \swift rms spectra, which may indicate a lack of significant \NH variability on long-timescales.

To be able to study the obscuration phenomenon and its variability in \ngc, it is essential to separate the variability of the intrinsic continuum components from that of the obscurer as both can produce similar X-ray variability. In the soft X-ray band, the continuum of \ngc is composed of an underlying power-law component ({\tt pow}) with the addition of a soft X-ray excess component ({\tt comt}). Thus, the variability of these two continuum components must be disentangled from that of the obscurer. This is achieved by broadband (optical-UV-X-ray) spectral modelling as performed in this paper. Since the soft excess component in \ngc is consistent with being the high-energy tail of the optical-UV continuum as explained by warm Comptonisation (see \citetalias{Meh14a}), the optical-UV data of \swift UVOT and HST COS enable us to constrain the strength of the soft excess emission. Thus, we can separate emission variability of the soft excess from absorption variability of the obscurer. Also, the 2--10 keV part of the \swift XRT spectrum allows us to constrain the power-law component, thus the variability of the power-law, the soft excess and the obscurer is determined.

The obscurer in \ngc, as well as heavily absorbing the soft X-rays, produces broad absorption lines (e.g. \ion{C}{iv}) in the UV band (see \citetalias{Kaas14}), while hardly absorbing the UV continuum. From spectroscopy of these UV lines, it is found that the obscurer is outflowing with velocities of up to $5000\ \kms$. Therefore, some of the observed changes in \cf of the obscurer are likely to be associated with its patchy nature and high outflow velocity. The obscurer, as well as having an outflow velocity component in our line of sight, is also likely to have a transverse velocity component. Therefore, the motion of the obscuring gas in our line of sight, plus any inhomogeneities within it, can result in changes in the covering of the underlying X-ray source. Since the partially-covering obscuration in \ngc has been persistent for at least 4 years and the \swift monitoring shows it has not yet cleared our line of sight since its discovery, this points to an outflowing structure with an elongated geometry, which most likely originates from the accretion disk given its close proximity to it. We note that there is currently an ongoing \swift monitoring of \ngc on a monthly basis (ending June 2016) and these latest \swift data show that \ngc is still obscured to this date (February 2016).

\subsection{Nature of the optical-UV-X-ray continuum variability}
\label{discuss_continuum}

The optical-UV continuum of \ngc agrees with a single thermal Comptonised component, also producing the soft X-ray excess. This has been shown in \citetalias{Meh14a} from broadband spectral modelling of stacked 2013 \hst, \xmm and \nustar spectra, as well as archival \xmm observations. This explanation is consistent with the results of our broadband spectral modelling of the \swift and HST COS data in this paper. In Fig. \ref{rms_spectra} we showed that the shape of the observed rms spectrum variability, as seen by the \swift UVOT and HST COS, is consistent with the model rms spectrum variability produced by changes in the normalisation and $T_{\rm seed}$ of the warm Comptonisation ({\tt comt}) component. The variability of {\tt comt} can explain the optical-UV variability at all times in \ngc. However, in Fig. \ref{rms_spectra} there are small differences between the observed and model rms variability in the optical-UV band. This is most likely due to variability of the strong AGN emission lines, including \ion{Fe}{ii} and the Balmer continuum, that fall in this band, and which their variability in the UVOT photometric filters were modelled as being only constant in our study. The emission from the BLR is known to be variable in \ngc, and in AGN in general, which is the basis of reverberation-mapping techniques \citep{Pet04,Cack15}. In our model, only optical-UV rms variability due to the variability of the continuum is considered, whereas the observed rms variability over a period of few months also includes variability of the BLR emission.

Apart from variability of the {\tt comt} component, there appears to be a long-term trend of power-law $\Gamma$ variability in \ngc (Figs. \ref{par_lc} and \ref{pow_gam_norm}), with $\Gamma$ being significantly lower in 2013 than in 2012 and 2014. We note that this hardening of the power-law photon index in \ngc does not have any apparent relation to the obscuration event. Low $\Gamma$ values (${\sim1.5}$) are found in data before \ngc became obscured: e.g. the 2005 and 2007 \swift data (Fig. \ref{pow_gam_norm}) and the 2000 \chandra data \citep{Yaqo01}.

Figure \ref{pow_gam_norm} shows that the photon index of the power-law becomes softer with increasing X-ray flux. We note that this behaviour was also found on shorter timescales by \citet{DiGesu14b}, as well as in previous X-ray monitorings of \ngc, such as with RXTE and ASCA by \citet{Chi00}. This brightening of the power-law as it gets softer is also found in similar Seyfert~1 AGN (e.g. \object{NGC 7469}, \citealt{Blu03}; \object{NGC 3516}, \citealt{Meh10}), and most likely originates in changes in the Compton up-scattering by the hot, optically thin corona which produces the primary power-law emission \citep{Haar93}. Basically, as the photon flux increases, more photons are scattered in the corona and so the corona loses energy, hence the steeper power-law. The positive correlation between the variability of the UV flux and the power-law normalisation and $\Gamma$ (found in Sect. \ref{para_results}) is also likely to be associated with the hot Comptonisation process. We note that in Fig. \ref{pow_gam_norm}, there is a possible indication that the power-law photon index stays unchanged at $\Gamma$ of about 2 for power-law normalisations $N_{\rm POW}$ above $1 \times 10^{52}$~photons~s$^{-1}$ keV$^{-1}$ at 1 keV. Such saturation of $\Gamma$ has previously been interpreted to be a signature of bulk-motion Comptonisation flow onto the black hole by \citet{Shapo09} and \citet{Tita16} in Galactic Black Hole Binaries.

Finally, in our \swift study of \ngc, we find that as the strength of the optical-UV/soft X-ray excess ({\tt comt}) component increases during periods of UV flaring, the covering fraction \cf of the obscurer decreases. We note that in our modelling, only the normalisation and $T_{\rm seed}$ of {\tt comt}, and \cf of the obscurer, were needed to fit the \swift and HST COS optical-UV/soft X-ray data. Thus, the impact of any smaller variation of other parameters, which were kept fixed in our modelling, is not known. However, in the analysis of \xmm and \chandra data by \citet{DiGesu14b}, where both \cf and \NH of the obscurer were fitted, similar correlations to ours were found between the parameters. The limitations of the \swift data and the modelling are unlikely to be responsible for producing erroneous correlations as similar results are obtained across our papers studying different aspects of \ngc. Although in the soft X-ray band both {\tt comt} and \cf can contribute to the rms spectral variability (as shown in Fig. \ref{rms_spectra}), the strength of the {\tt comt} component is primarily determined from the optical-UV flux. Note that the optical-UV continuum below the Lyman limit (13.6 eV) is unaffected by the obscuration. With simultaneous optical-UV data from UVOT and HST COS, it is possible to disentangle changes in the strength of the soft excess flux from the covering fraction \cf of the obscurer. Thus, the obtained anti-correlation is unlikely to be a mere artefact of spectral fitting. One possible explanation for this apparent anti-correlation between the covering fraction of the obscurer and the strength of the soft excess emission during UV flaring could be if the size of the soft excess emitting coronal region in the inner regions of the disk varies. In this picture, as the area emitting the soft excess becomes larger (i.e. $r_{\rm{corona}}$ becomes larger), the nearby obscurer in our line of sight covers less of the X-ray emitting region. Hence, the observed \cf becomes smaller. The details of the corresponding decrease in \cf depend on the uncertain geometry of the stream of obscuring gas relative to the corona in our line of sight. However, in simple terms the new covering fraction would be approximately ${C^{\,\prime}_f \propto C_f\, (r_{\rm{corona}} / r^{\,\prime}_{\rm{corona}})^2}$.

\section{Conclusions}
\label{conclusions}

We have studied the optical-UV-X-ray variability of Seyfert~1 galaxy \ngc using remarkably extensive and intense monitoring campaigns with \swift and HST COS. By performing broadband modelling of the spectra and rms variability analysis of the lightcurves, we have disentangled intrinsic continuum variability from absorption variability of the obscuring outflow near the disk. From the findings of our investigation we conclude that:
\begin{enumerate}
\item The X-ray obscurer (which is composed of outflowing weakly-ionised gas extending from near the disk to the BLR) has been continuously present in our line of sight since at least February 2012. It has been partially covering the X-ray source since its discovery to this date, and has never cleared our line of sight. The \swift data show that \ngc was unobscured in August 2007. However, due to lack of any monitoring between August 2007 and February 2012, it is not possible to more precisely determine when the transition between the unobscured and obscured states occurred. \\

\item The covering fraction of the obscurer is time variable, ranging between ${0.70 \le \cf \le 0.98}$. The observed X-ray hardness ratio ($R$) variability is predominantly caused by changes in the covering fraction of the obscurer. Some of the \cf changes are likely to be due to the transverse motion of the outflowing obscurer in our line of sight and its patchy nature. The persistence of such partially-covering and outflowing obscuration over several years indicates an outflowing structure with an elongated geometry. \\

\item Our analysis of the \swift data confirms that the soft X-ray excess in \ngc is consistent with being the high-energy tail of the optical-UV continuum, which can be explained by warm Comptonisation: up-scattering of the disk seed photons in a warm, optically thick corona as part of the inner disk. \\

\item The \swift data reveal that when the optical-UV flux flares up (steep continuous increase over 1--2 months), the hardness ratio $R$ and the obscurer \cf become smaller. This may imply that as the optical-UV/soft excess emitting warm corona on the surface of the disk becomes larger during these periods, the obscurer stream covers less of the X-ray source.
\end{enumerate}

\begin{acknowledgements}

This work made use of data supplied by the UK Swift Science Data Centre at the University of Leicester. SRON is supported financially by NWO, the Netherlands Organization for Scientific Research. M.M. acknowledges support from NWO and the UK STFC. This work was supported by NASA through grants for HST program number 13184 from the Space Telescope Science Institute, which is operated by the Association of Universities for Research in Astronomy, Incorporated, under NASA contract NAS5-26555. M.C. acknowledges financial support from contracts ASI/INAF n.I/037/12/0 and PRIN INAF 2011 and 2012. P.-O.P. acknowledges financial support from the CNES and from the CNRS/PICS. G.P. acknowledges support via an EU Marie Curie Intra-European fellowship under contract no. FP-PEOPLE-2012-IEF-331095 and Bundesministerium f\"{u}r Wirtschaft und Technologie/Deutsches Zentrum f\"{u}r Luft- und Raumfahrt (BMWI/DLR, FKZ 50 OR 1408). K.C.S. acknowledges financial support from the Fondo Fortalecimiento de la Productividad Cient\'{i}fica VRIDT 2013. E.B. received funding from the EU Horizon 2020 research and innovation programme under the Marie Sklodowska-Curie grant agreement no. 655324, and from the I-CORE program of the Planning and Budgeting Committee (grant number 1937/12). S.B. and G.M. acknowledge INAF/PICS financial support. G.M. and F.U. acknowledge financial support from the Italian Space Agency under grant ASI/INAF I/037/12/0-011/13. B.M.P. acknowledges support from the US NSF through grant AST-1008882. F.U. acknowledges PhD funding from the VINCI program of the French-Italian University. M.W. acknowledges the support of a PhD studentship awarded by the UK Science \& Technology Facilities Council (STFC). We thank the International Space Science Institute (ISSI) in Bern for their support and hospitality. We thank the anonymous referee for her/his comments.

\end{acknowledgements}



\appendix

\section{Relations between the optical and UV lightcurves of \ngc}
\label{uv_appendix}

\begin{figure}[!tbp]
\centering
\resizebox{0.85\hsize}{!}{\includegraphics[angle=0]{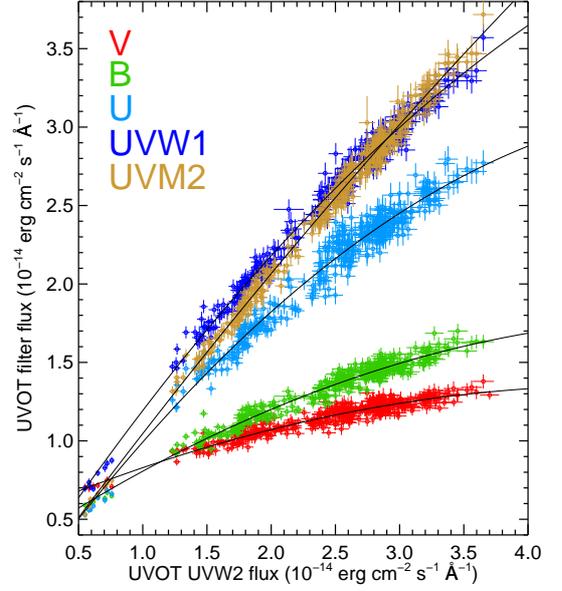}}
\caption{Relations between the observed optical and UV fluxes of \ngc taken with the six photometric filters of \swift UVOT. The fitted functions are given in Eqs. \ref{uvot_rel_best_eqs}.}
\label{uvot_relation}
\end{figure}

Figure \ref{uvot_relation} shows how the optical and UV fluxes taken in the \swift UVOT photometric filters are related to each other. From fitting the flux in different filters versus the flux in the UVW2 filter, we derive the following empirical relations (in units of $10^{-14}$ \ergcm) described by quadratic functions:
\begin{subequations}
\label{uvot_rel_best_eqs}
\begin{eqnarray}
F_{\rm{V}} = 0.512  + (0.354\ F_{\rm{UVW2}}) - (0.037\ {F_{\rm{UVW2}}}^2)  \\
F_{\rm{B}} = 0.312 + (0.545\ F_{\rm{UVW2}}) - (0.050\ {F_{\rm{UVW2}}}^2)  \\
F_{\rm{U}} = -0.039 + (1.132\ F_{\rm{UVW2}}) - (0.101\ {F_{\rm{UVW2}}}^2)  \\
F_{\rm{UVW1}} = 0.039 + (1.238\ F_{\rm{UVW2}}) - (0.084\ {F_{\rm{UVW2}}}^2)  \\
F_{\rm{UVM2}} = -0.042 + (1.123\ F_{\rm{UVW2}}) - (0.034\ {F_{\rm{UVW2}}}^2)
\end{eqnarray}
\end{subequations}

We note that the UVOT flux values in the above relations are the observed flux and do not necessarily represent the flux of the continuum. They include flux contributions from the BLR, the NLR and the host galaxy stellar emission, as well as Galactic reddening, which are all taken into account separately during the spectral fitting in {\tt SPEX} using the optical/UV correcting components reported in \citetalias{Meh14a}. The above empirical relations have been used to calculate the observed UVOT flux for those \swift observations when exposures in some of the six UVOT filters have not been taken. This method is useful in the context of broadband spectral modelling of the \swift data, as it provides us with the same number of data sets (covering identical energy bands) in all the \swift observations, in order to carry out a uniform and consistent spectral fitting of the data.

\begin{figure}[!tbp]
\centering
\resizebox{0.85\hsize}{!}{\includegraphics[angle=0]{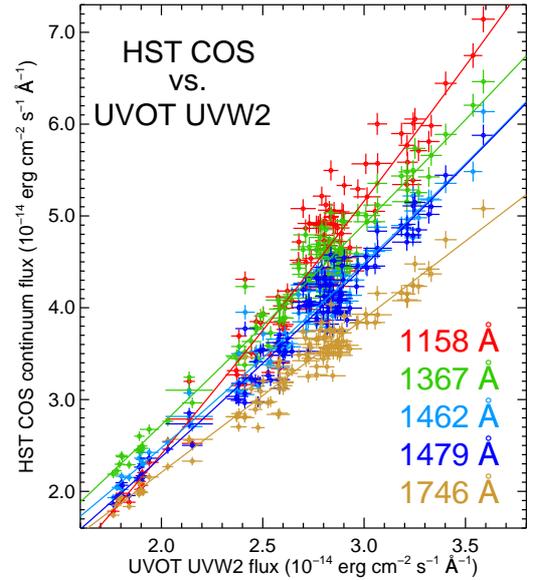}}
\caption{HST COS continuum fluxes of \ngc plotted versus contemporaneous UVOT UVW2 flux. The observed COS fluxes are taken from five narrow energy bands, which are free of emission and absorption features to represent the continuum. The fitted functions are given in Eqs. \ref{cos_rel_best_eqs}.}
\label{uvot_cos_relation}
\end{figure}

In Fig. \ref{uvot_cos_relation} we show the flux relations between the contemporaneous \hst COS and UVOT UVW2 observations (separated by less than 24 hours). From fitting the COS continuum flux at different wavelengths versus the UVW2 filter flux, we derive the following empirical relations (in units of $10^{-14}$ \ergcm) described by quadratic functions: 
\begin{subequations}
\label{cos_rel_best_eqs}
\begin{eqnarray}
F_{{1158\ \AA}} = -2.249  + (1.992\ F_{\rm{UVW2}}) + (0.155\ {F_{\rm{UVW2}}}^2)  \\
F_{{1367\ \AA}} = -1.235 + (1.811\ F_{\rm{UVW2}}) + (0.076\ {F_{\rm{UVW2}}}^2)  \\
F_{{1462\ \AA}}  = -0.649 + (1.235\ F_{\rm{UVW2}}) + (0.157\ {F_{\rm{UVW2}}}^2)  \\
F_{{1479\ \AA}} = -1.464 + (1.770\ F_{\rm{UVW2}}) + (0.067\ {F_{\rm{UVW2}}}^2)  \\
F_{{1746\ \AA}} = -1.072 + (1.576\ F_{\rm{UVW2}}) + (0.025\ {F_{\rm{UVW2}}}^2)
\end{eqnarray}
\end{subequations}

The observed COS fluxes in the above relations include the Galactic reddening and are from five narrow energy bands, which are free of emission and absorption features to represent the continuum. The above relations were used to calculate the COS continuum fluxes for those \swift observations without contemporaneous \hst COS data. These predicted COS fluxes were used in our broadband spectral modelling in order to constrain the far-UV continuum.

%

%

\end{document}